\documentclass[a4paper,12pt]{article}
\pdfoutput=1

\usepackage[bottom=26mm, width=18cm]{geometry}
\selectfont

\usepackage{graphicx}
\usepackage{amssymb}
\usepackage{amsmath,amssymb,amsthm,bm}
\usepackage[utf8]{inputenc}
\usepackage{tabularx}
\usepackage{geometry,graphicx}
\newtheorem{remark}{Remark}

\newtheorem{result}{Result}
\newtheorem{definition}{Definition}
\usepackage{bigints} 
\usepackage{natbib}
\usepackage{caption}
\newcommand{\ind}{\mbox{$\perp\!\!\!\perp \,$}}

\theoremstyle{plain}

\theoremstyle{definition}

\begin{document}

\title{Graphical models for circular variables}

\author{Anna Gottard and Agnese Panzera \\
\\ Department of Statistics, Computer Science, Applications\\ University of Florence, Italy}

\maketitle
\abstract{
Graphical models are a key class of probabilistic models for studying the conditional independence structure of a set of random variables.
Circular variables are special variables, characterized by periodicity, arising in several contexts and fields. However, models for studying the dependence/independence structure of circular variables are under-explored.
This paper analyses three multivariate circular distributions, the von Mises, the Wrapped Normal and the Inverse Stereographic distributions, focusing on their properties concerning conditional independence.   
For each one of these distributions, we discuss the main properties related to conditional independence and introduce suitable classes of graphical models. The usefulness of the proposed models is shown by modelling the conditional independence among dihedral angles characterizing the three-dimensional structure of some proteins.  
}


\section{Introduction}
A circular observation can be regarded as a point on the circumference of the unit circle, as a unit vector on the plane or as a unit complex number. Once both an origin and a sense of rotation have been chosen, a circular observation can be measured by an angle which, in radians, ranges from $0$ to $2\pi$ or, equivalently, from $-\pi$ to $\pi$. Typical examples of circular data include directions of migration of the birds from the point of release, directions of the winds and marine currents, the angles in the polypeptide chains forming proteins, but also the time of the day at which a given event occurs. 
The special nature of circular data lies in their \emph{periodicity}, i.e. a circular observation measured by an angle of $\theta$ radians corresponds, on the unit circle,  
to a circular observation measured by an angle of $\theta+2\pi k $ radians, where $k$ is a whatever integer number.

\emph{Circular statistics} collects ad hoc statistical methods for dealing with circular data. Probability distributions of circular random variables can be defined according to three main approaches. The \emph{embedding} approach obtains circular distributions as radial projections onto the unit circle of distributions on the real plane.  The \emph{intrinsic} approach identifies the sample space as the circle itself with probability distributions directly defined on it.  Finally, the \emph{wrapping} approach constructs circular random variables as the modulo $2\pi$ version of the real ones with distributions obtained by wrapping their real counterparts around the circle.
A comprehensive account on circular statistics and circular distributions is provided, among others, by \cite{jammalamadaka2001topics}, and \cite{mardia2009directional}. Some recent advances are collected by \cite{ley2017modern}, and \cite{ley2018applied}.

In some applications, the focus is on several angles, say $p$, which can be regarded as points on the surface of a $p-$dimensional torus, $\mathbb{T}^p$, obtained by the $p-$fold Cartesian product of unit circles. A typical example of toroidal data arises in the three-dimensional protein structure, which can be summarised by a sequence of angles.
Despite the advances in circular statistics, both distributions on $\mathbb{T}^p$ and models involving $p$ circular variables seem to be not deeply explored in the literature whenever $p>2$. 

 A topic that is arousing some interest is the study of the dependence and conditional independence structure of a vector of circular variables. Graphical models \citep[see, among many,][]{lauritzen1996} are a powerful probabilistic tool for analysing conditional independences between random variables.
This class of multivariate models expresses the conditional independence structure of $p$ variables by a graph $G=(V, E)$. Here random variables are represented by the nodes of the graph, collected in the set $V = \{1, \ldots, p \}$, and the graph topology displays the conditional independence statements as missing edges in the set $E\subset V\times V$. A graphical model is called \emph{undirected graph model} or \emph{Markov network} whenever the associated graph $G$ is undirected, that is when $E\subseteq \{\{i,j\}:i,j\in V,i\neq j\}$  is a set of unordered pairs. This model considers variables all on equal footing. The most used undirected graphical model is the \emph{concentration graph model} or \emph{Gaussian graphical model} that assumes a joint Gaussian distribution for the $p$ variables. For these models,  learning the missing edges is equivalent to identifying the zero elements in the inverse of the covariance matrix. 
Conversely, a graphical model is called \emph{Bayesian network} when $G$ is a directed acyclic graph having only directed edges, that is $E\subseteq \{(i,j): i,j\in V,i\neq j\}$ is a set of ordered edges, but with no cycles allowed.  Bayesian networks assume a complete ordering of the studied variables and provide a useful factorisation in terms of univariate conditional distributions related to the chain rule of probability. 

Even if graphical models for circular variables can be extremely useful in many research areas, except for some relatively recent contributions, this topic seems to be under-explored. The first proposal, up-to our knowledge, is provided by \cite{boomsma2006graphical}. They studied the joint distribution over a protein sequence and structure, by using two classes of models which combine Hidden Markov Models, as Bayesian networks, with distributions which are respectively defined on the sphere and on the bidimensional torus.   
Later, \cite{razavian2011mises} studied conditional independence among angles having the sine version of a $p-$variate von Mises distribution and represented it via a factor graph. As this distribution does not admit a closed-form of the normalising constant for $p>2$, the Authors adopted the full pseudo-likelihood approach for inference \citep{mardia2008} based on the property that the univariate full conditional distributions are von Mises. Nevertheless, as we will see, the lack of closeness under marginalisation and conditioning of this distribution limits its possible use as a graphical model. Recently, \cite{leguey2016tree} considers the bivariate wrapped Cauchy distribution which can be associated with a tree-structured directed acyclic graph. The considered distribution, being defined on $\mathbb{T}^2$, does not allow the extension to graphs without tree structure.  \cite{klein2019torus} introduced a class of graphical models for data lying on $\mathbb{T}^p$, called \emph{torus graphs}, to tackle the problem of identifying phase coupling among oscillatory signals recorded from multiple electrodes in the brain. These models are defined not on a specific distribution but as members of a full exponential family with pairwise interactions.  In particular,  \cite{klein2019torus} focused on three subfamilies of torus graphs which are all especially useful for their motivating application. These subfamilies are the uniform marginal model, the phase difference model, and a phase difference model with uniform margins.   

In this paper, we aim to explore in terms of conditional independence and graphical models three distributions on $\mathbb{T}^p$ involving the three main approaches for circular distributions.
Regarding the intrinsic approach, we further discuss the sine version of the von Mises distribution.  In particular, after pointing out some issues of this distribution as an undirected graph model, we propose a related class of directed acyclic graph models when the ordering of the angles is known \emph{a priori}. These models can be useful, for instance, to study the sequence of the dihedral angles assuming as given the primary structure of a protein, or to study the trajectory of an animal regarded as a sequence of directions. For both the wrapped and the embedded approaches, we introduce two classes of undirected graphical models, which are related, in different manners, to the \emph{classical} Gaussian graphical model. Specifically, the first class is based on the Wrapped Normal distribution, while the second is based on the inverse stereographic projected Normal distribution, as considered in \cite{selvitella2019geometric}. For both these classes of undirected graphical models, we discuss issues and some possible solutions, also including a semi-parametric model.

The paper is organised as follows.
Section \ref{vM} is devoted to discussing graphical models for the von Mises distribution. Section \ref{WN} presents some properties of the Wrapped Normal related distributions and a subsequent class of graphical models. 
Section \ref{ISN} introduces a class of graphical models for the inverse stereographic distribution and a related semi-parametric version. Section \ref{ex} contains some illustrative examples where the introduced models are employed to describe the relationships among angles in the structure of some proteins. Section \ref{concl} concludes with some final considerations.

\section{von Mises related graphical models} \label{vM}
A circular random variable has a von Mises distribution with \emph{mean direction} $\mu\in\mathbb{T}$, and \emph{concentration parameter} $\kappa\geq0$ if its probability density function, at $\theta\in\mathbb{T}$, is

$$f_{\Theta}(\theta)=\frac{1}{2\pi\mathcal{I}_0(\kappa)}\exp\left(\kappa\cos(\theta-\mu)\right),$$ 
where $\mathcal{I}_0(\kappa)$ stands for the modified Bessel function of first kind and order $0$. 

The multivariate von Mises distribution on $\mathbb{T}^p$, as firstly introduced in \cite{mardia2008}, is also known as the multivariate sine distribution.
Let $\bm \mu\in\mathbb{T}^p$, $\bm \kappa\in\mathbb{R}^p$, with $\kappa_i\geq 0$,  and $\bm \Lambda\in\textsf{Sym}^{p\times p}$, with $\lambda_{ii}=0$ for $i\in\{1,\dots,p\}$. A vector of random angles $\bm{\Theta}=(\Theta_1\dots\Theta_p)'$ has a $p-$variate sine distribution, $\bm \Theta\sim vM_p(\bm \mu,\bm{\kappa},\bm \Lambda)$,  if its probability density function at $\bm\theta\in\mathbb{T}^p$ is
$$f_{\bm \Theta}(\bm \theta)=\frac{1}{C(\bm \kappa,\bm\Lambda)}\exp\left(\bm \kappa'c(\bm \theta,\bm{\mu})+\frac{1}{2}s(\bm \theta,\bm{\mu})'\bm{\Lambda}s(\bm{\theta},\bm{\mu})\right),$$ where $C(\bm \kappa,\bm\Lambda)$ is a normalization factor and $c(\bm\theta,\bm{\mu})$ and $s(\bm \theta,\bm{\mu})$ are $p-$dimensional vectors  having $\cos(\theta_i-\mu_i)$ and $\sin(\theta_i-\mu_i)$ as their respective $i^\text{th}$ entry. 
The factor $C(\bm \kappa,\bm\Lambda)$ has no closed form when $p>2$.

If  $\bm \Theta\sim vM_p(\bm \mu,\bm{\kappa},\bm \Lambda)$, except for the case of independent random angles, the marginal distributions are not von Mises,  while the  univariate conditional distributions are von Mises. In particular,  the  distribution of $\Theta_j$ given $\bm{\Theta}_\text{rest}$, with $\text{rest}=\{1,\dots,j-1,j+1,\dots,p\}$ is

$$f_{\Theta_j\mid\bm \Theta_\text{rest}}(\theta_j\mid \bm\theta_\text{rest})=\frac{ \exp(\kappa_{j.\text{rest}}\cos(\theta_j-\mu_{j.\text{rest}}))}{2\pi\mathcal{I}_0(\kappa_{j.\text{rest}})},$$
where
$$\tan(\mu_{j.\text{rest}})=\mu_j+\sum_{i\neq j} \lambda_{ij}\sin(\theta_i)/\kappa_j,$$
and
$$\kappa_{j.\text{rest}}=\left\{\kappa_j^2+\left(\sum_{i\neq j}\lambda_{ij}\sin(\theta_i)\right)\right\}^{1/2}.$$
Then, $\Theta_j\ind \Theta_i\mid \bm \Theta_{\text{rest}\setminus i}$ iff $\lambda_{ij}=0$. 
However,   the conditional distributions of $\bm \Theta_{\bm r}$, for a subset $\bm r$ of $ \{1,\ldots, p\}$ with  cardinality greater than $1$, are not von Mises. Summarizing, the sine distribution is not closed either under marginalisation or under conditioning and $\lambda_{ij}$
plays the role of the pairwise conditional independence parameter.

As aforementioned, to the best of our knowledge, there are only two contributions where the sine distribution has been contemplated in the framework of graphical models. \cite{razavian2011mises}, thanks to the property of the sine distribution to involve only pairwise interactions, define a \emph{factor graph} representation of this distribution. The resulting graph obeys to the pairwise Markov property \citep{lauritzen1996}.  
Typically,  graphical models defined for undirected graphs are closed both under marginalisation and conditioning \citep[see, for instance,][]{richardson2002, cox2014}, so that one can derive from the graph $G$ of the joint distribution, the graph representing conditional independences holding in any marginal and conditional distribution. Unfortunately, as mentioned before, the von Mises is not closed under marginalisation and conditioning. 
However, as this distribution is strictly positive, the global Markov property is implied by the pairwise for the Hammersley-Clifford Theorem \citep{lauritzen1996}. In addition, according to  \cite{sadeghi2013},   all the marginal and conditional distributions, whatever they are, are also Markov to the graph $G$. Nevertheless, due to the lack of closeness, some desirable properties such as decomposability and collapsibility cannot be used in  \cite{razavian2011mises}'s graphical model.

As the second contribution, \cite{klein2019torus} mentioned the sine model as a restriction of the natural parameter space of the torus graph density. Nonetheless, while the torus graph is defined as a regular full exponential family closed under marginalisation and conditioning,  the sine distribution is not a regular exponential family. Therefore, the sine distribution is not a torus graph model.

As a further class of von Mises-based graphical models,  we propose a class of directed acyclic graphical models, suitable when the ordering of the circular random variables is known. These models are defined as a sequence of random angles having univariate conditional von Mises distributions. In particular, letting  $\text{pa}(j)$ be the set of parents of the vertex $j\in \{1,\dots,p\}$, that is the set of vertices such that there is an arrow from them to $j$, we get the following definition.
\begin{definition}\label{condvM} \textbf{\emph{(Conditional von Mises directed acyclic graphical  model)}} 	Let $\mathcal{G}=(V,E)$ be a directed acyclic graph with set of ordered vertices $V=\{1,\ldots,p\}$ and set of directed edges $E\subset V\times V$. 
	We say that  $\bm{\Theta}=(\Theta_1\ldots\Theta_p)'$ is a conditional von Mises directed acyclic graphical  model with respect to  $\mathcal{G}$ if 
	the probability distribution of $\bm{\Theta}$ satisfies

	$$f_{\bm{\Theta}}(\bm \theta) = \prod_{j=1}^{p} f_{\Theta_j\mid \bm\Theta_{\mathrm{pa}(j)}}(\theta_j |\bm\theta_{\mathrm{pa}(j)}), $$
	
	where  
	
	$$f_{\Theta_j\mid  \bm\Theta_{\mathrm{pa}(j) }}(\theta_j|\bm\theta_{\mathrm{pa}(j)})=\frac{ \exp\{\kappa_{j.\mathrm{pa}(j)}\cos(\theta_j-\mu_{j.\mathrm{pa}(j)})\}}{2\pi\mathcal{I}_0(\kappa_{j.\mathrm{pa}(j)})},$$
	where
	$$\tan(\mu_{j.\mathrm{pa}(j)})=\mu_j+\tan^{-1}\left(\sum_{i\in \mathrm{pa}(j)} \lambda_{ij}\sin(\theta_i)/\kappa_j\right),$$
	and
	$$\kappa_{j.\mathrm{pa}(j)}=\left\{\kappa_j^2+\left(\sum_{ i\in \mathrm{ pa}(j)}\lambda_{ij}\sin(\theta_i)\right)\right\}^{1/2}.$$
For each $i<j$,  
$$i \notin \mathrm{pa}(j)\;\; \text{and}\;\; (i,j)\notin E \iff \lambda_{ij} = 0.$$
\end{definition}

Notice that the distribution of $\bm \Theta$ is not von Mises but it is the product of univariate von Mises. For this reason, while the set of parents of each node can be learned from data, the node ordering has to be known a priori and standard structure learning algorithms such as the PC-Algorithm \citep{spirtes2000} cannot be used. 
These models can be useful, for example, to study the sequence of the dihedral angles in a protein when the ordering of the angles is provided by the primary structure of the protein itself.

Assuming univariate conditional von Mises allows a model where both the mean direction and the concentration depend on the set of parents of each node. Conversely, one can assume only the mean direction to be dependent on the parents, adopting, for instance, the specification of conditional von Mises proposed by \cite{fisher1992regression}.

An interesting aspect of the conditional von Mises directed acyclic graphical model is that it can be easily extended to the case of mixed-type variables, where some variables are circular, and others are not.  In such a situation, the conditional distribution of the non-circular variable has to be replaced in the factorisation with another conditional distribution suggested by the nature of the variable itself. These models could be useful, for instance, to study a sequence of animal movements, together with the time of the day, the weather and the temperature.

\section{Wrapped Normal related graphical models}\label{WN}
According to the wrapping approach, any linear random variable $X$ can be transformed into a circular variable $\Theta$ by reducing its modulo $2\pi$, i.e.
$$\Theta=X(\mathrm{mod} 2\pi).$$
This approach corresponds to wrapping the real line around the unit circle, accumulating probability over all overlapping points $\theta$, $\theta\pm 2\pi$, $\theta\pm4\pi$, and so on. Denoting as $f_X$ the probability density function  of $X$, the corresponding probability density function of $\Theta$, at $\theta\in\mathbb{T}$, is defined as
$$f_{\Theta}(\theta)=\sum_{k=-\infty}^\infty f_X(\theta+2\pi k).$$
When $X\sim N(\mu,\sigma^2)$, then $f_{\Theta}$ is the probability density function of a Wrapped Normal distribution.

In the multivariate setting, a vector of random angles $\bm \Theta=(\Theta_1\ldots\Theta_p)'$ has a Wrapped Normal (WN) distribution, $\bm \Theta\sim WN_p(\bm{\mu},\bm{\Sigma})$,  with parameters $\bm \mu\in\mathbb{R}^p$ and $\bm{\Sigma}\in  \textrm{Sym}_+^{p\times p}$, if $$\bm X=\bm{\Theta}+2\pi\bm K$$ is such that $\bm X\sim N_p(\bm \mu,\bm{\Sigma})$ and $\bm K$ is a \emph{latent} random vector taking values on $\mathbb{Z}^p$. Hence $\bm X$ determines both $\bm\Theta$ and $\bm K$ via modulo operation. For further details on this approach to define the wrapped Normal distribution, see \cite{jona2012}.  Denoting by $f_{\bm X}$ the probability density function of $\bm X$, the joint density of  $(\bm\Theta,\bm K)$ is again $f_{\bm X}$, while  the  density of $\bm\Theta$, at $\bm{\theta}\in\mathbb{T}^p$, is 
\[f_{\bm{\bm\Theta}}(\bm{\theta})=\sum_{\bm k\in\mathbb{Z}^p}f_{\bm X}(\bm\theta+2\pi\bm{k}).\]

The following remark provides a key for the interpretation of the parameters of a WN distribution for the complex form representation of $\bm{\Theta}$.
\begin{remark}\label{remarkcomplex}
	Consider the complex form representation of $\bm \Theta$, that is $\mathbf{Z}=(Z_1\ldots Z_p)'$, where $Z_j=e^{\mathrm i\Theta_j}$, with $\mathrm{i}^2=-1$ and $j\in\{1,\ldots, p\}$. Now, if $\bm{\Theta}\sim WN_p(\bm \mu,\bm{\Sigma})$, then
	$$\mathbb{E}[ Z_j]=e^{-(\bm\Sigma)_{jj}/2}e^{\mathrm i\mu_j}=e^{-(\bm\Sigma)_{jj}/2}\{\cos(\mu_j)+\mathrm{i} \sin(\mu_j)\}.$$
	Thus the mean resultant length  $\Vert\mathbb{E}[Z_j]\Vert=e^{-(\bm\Sigma)_{jj}/2}$ plays the role of the concentration parameter for $\Theta_j$. Moreover, as $\Theta_i+\Theta_j=(X_i+X_j)\mathrm{mod} (2\pi)$,
	$$\mathbb{E}[Z_iZ_j]=e^{-(\bm\Sigma)_{ii}/2}e^{-(\bm\Sigma)_{jj}/2}e^{-(\bm \Sigma)_{ij}}e^{\mathrm i(\mu_i+\mu_j)}.$$
	In addition, as $\Vert\mathbb{E}[Z_i]\mathbb{E}[Z_j]\Vert=e^{-[(\bm\Sigma)_{ii}+(\bm\Sigma)_{jj}]/2},$ we have 
	$$\Vert\mathbb{E}[Z_iZ_j]\Vert= \Vert\mathbb{E}[Z_i]\mathbb{E}[Z_j]\Vert e^{-(\bm\Sigma)_{ij}},$$
	which yields
	$$e^{-(\bm\Sigma)_{ij}}=\frac{\Vert\mathbb{E}[Z_iZ_j]\Vert}{\Vert\mathbb{E}[Z_i]\mathbb{E}[Z_j]\Vert}.$$
	Then, the mean independence of $Z_i$ and $Z_j$ is equivalent to the independence in probability of $X_i$ and $X_j$.
\end{remark}
Now, let us consider some issues related to independence and conditional independence for wrapped Normal random angles.

Given  $\bm \Theta\sim WN_p(\bm\mu,\bm{\Sigma})$, consider the  partition of $\bm \Theta$ into  the components $\bm \Theta_A$ and $\bm \Theta_B$ which, for integer $q<p$, respectively take values on $\mathbb{T}^q$ and $\mathbb{T}^{p-q}$, and let $\bm \mu$ and $\bm{\Sigma}$ be partitioned accordingly, i.e.

\begin{equation}\label{partition}\bm \Theta=\left(\begin{array}{c}\bm{\Theta}_A\\\bm{\Theta}_B\end{array}\right),\;\;\;\bm \mu=\left(\begin{array}{c}\bm \mu_A\\\bm \mu_B\end{array}\right),\;\;\;\bm \Sigma=\left(\begin{array}{cc}\bm{\Sigma}_{AA}&\bm{\Sigma}_{AB}\\\bm{\Sigma}_{BA}&\bm{\Sigma}_{BB}\end{array}\right),\end{equation}
then, the following results hold.

\begin{result}  \label{marginal}
	Given the  vector of random angles $\bm \Theta\sim WN_p(\bm \mu,\bm{\Sigma})$, for any non-empty subset $A$ of  $\{1, \ldots , p\}$ of cardinality $q<p$,

	$$\bm\Theta_A \sim WN_q (\bm\mu_A, \bm \Sigma_{AA}).$$
	
\end{result}
\begin{proof}See Appendix.
\end{proof}



\begin{result} \label{conditionalX}
	Given the vector of random angles $\bm \Theta\sim WN_p(\bm \mu,\bm{\Sigma})$, for disjoint non-empty subsets $A$ and $B$ of $\{1,\dots,p\}$ with respective cardinalities $q<p$ and $p-q$ 
	
	$$\bm\Theta_A\mid \bm\Theta_B,\bm K_B\sim WN_q(\bm \mu_{A\mid B}, \bm{\Sigma}_{A\mid B}),$$
	where $\bm K_B$ is such that $\bm X_B=\bm{\Theta}_B+2\pi\bm K_B$, with $\bm X_B\sim N_{p-q}(\bm \mu_B,\bm{\Sigma}_B)$,
	and 
	$$\bm \mu_{A\mid B}=\bm{\mu}_A-\bm{\Sigma}_{AB}\bm{\Sigma}^{-1}_{BB}(\bm x_B-\bm{\mu}_B),\;\;\;\text{and}\;\;\;\;\bm{\Sigma}_{A\mid B}=\bm{\Sigma}_{AA}-\bm{\Sigma}_{AB}\bm\Sigma_{BB}^{-1}\bm{\Sigma}_{BA}.$$
	Moreover, $\bm \Theta_A\ind \bm{\Theta}_B$ iff $\bm \Sigma_{AB}$ is the null matrix $\bm 0_{q \times (p-q)}$.
\end{result}

\begin{proof} See Appendix.  
\end{proof}

Recalling Remark \ref{remarkcomplex}, the above result implies that  if $\bm \Theta\sim WN_p(\bm \mu,\bm{\Sigma})$ and $\bm Z=e^{-\mathrm{i}\bm{\Theta}}$, the  mean independence of $Z_i$ and $Z_j$ is equivalent to the independence in probability of $\Theta_i$ and $\Theta_j$.

\begin{result}\label{conditional}
	
	Given the  vector of random angles $\bm \Theta\sim WN_p(\bm \mu,\bm{\Sigma})$, the conditional distribution of $\bm\Theta_A\mid \bm\Theta_S$, with 
	$A, S$ being disjoint subsets of  $\{1, \ldots , p\}$,  with cardinality $q$ and $s$ respectively, $s \leq p-q$. Then, 
	
	$$
	f_{\bm\Theta_A\mid \bm\Theta_S}(\bm\theta_A\mid \bm\theta_S) = 
	\sum_{\bm k_A\in\mathbb{Z}^q}\sum_{\bm k_S\in\mathbb{Z}^s} 
	f_{\bm X_A\mid \bm X_S}(\bm\theta_A + 2 \pi\bm k_A \mid \bm\theta_S + 2 \pi\bm k_S) w_S(\bm \theta_S, \bm k_S)
	$$
	where 
	$$
	w_S(\bm \theta_S, \bm k_S) = \frac{f_{\bm X_S}( \bm\theta_S + 2 \pi\bm k_S) }{\sum_{\bm k_S\in\mathbb{Z}^s} f_{\bm X_S}( \bm\theta_S + 2 \pi\bm k_S)}.
	$$
\end{result}

\begin{proof} See Appendix.
\end{proof}


Notice that Result \ref{conditional}  shows that the conditional circular density is obtained by wrapping a mixture of the corresponding real distributions. Then, the resulting conditional distribution is not a wrapped Normal. This issue has been pointed out, in the different setting of toroidal diffusion processes, by \cite{garcia2019}. 

Now,
letting  $A$, $C$ and $S$ be non-empty  disjoint subsets of $\{1, \ldots, p\}$ with respective cardinalities $q$, $c$ and $s$ such that $q+c+s \leq p$,  then
\begin{align*}
	& f_{\bm\Theta_A\mid \bm\Theta_C,\bm\Theta_S}(\bm\theta_A\mid \bm\theta_C, \bm\theta_S)  = \\
	& = \sum_{\bm k_A\in\mathbb{Z}^q}\sum_{\bm k_C\in\mathbb{Z}^c}\sum_{\bm k_S\in\mathbb{Z}^s} 
	f_{\bm X_A\mid \bm X_C \bm X_S}(\bm\theta_A + 2 \pi\bm k_A \mid \bm\theta_C + 2 \pi\bm k_C, \bm\theta_S + 2 \pi\bm k_S)w_{CS}(\bm \theta_C,\bm k_C, \bm \theta_S,\bm k_S)
\end{align*}
with 
$$
w_{CS}(\bm \theta_C,\bm k_C, \bm \theta_S,\bm k_S) = \frac{f_{\bm X_C\bm X_S}( \bm\theta_C + 2 \pi\bm k_C, \bm\theta_S + 2 \pi\bm k_S) }{\sum_{\bm k_C\in\mathbb{Z}^c}\sum_{\bm k_S\in\mathbb{Z}^s} f_{\bm X_C\bm X_S}( \bm\theta_C + 2 \pi\bm k_C, \bm\theta_S + 2 \pi\bm k_S)}.
$$
Now if $\bm\Theta_A \ind \bm\Theta_C \mid \bm\Theta_S$, then we should have $f_{\bm\Theta_A\mid \bm\Theta_C,\bm\Theta_S}(\bm\theta_A\mid \bm\theta_C, \bm\theta_S) = f_{\bm\Theta_A\mid \bm\Theta_S}(\bm\theta_A\mid  \bm\theta_S)$. Unfortunately, because of  the sum over $\bm k_C$, there is no value or combination of values of the parameters ensuring conditional independence.

Therefore, from the point of view of graphical models, the main issue for the WN distribution is not the lack of closeness under conditioning, but rather the confounding due to the marginalisation over the vector of winding numbers. See, among many, \cite{wermuth2011summary} for the effect of marginalisation on undirected graph models.
For all these reasons, it seems hard to define this distribution as a graphical model. However, it is quite straightforward to define a graphical model for the unwrapped variables $\bm X=\bm{\Theta}+2\pi\bm K$ as follows.

\begin{definition}\label{unwrapped} \textbf{\emph{(Unwrapped Normal graphical model)}}
	Let $\mathcal{G}=(V,E)$ be an undirected graph with set of vertices $V=\{1,\ldots,p\}$ and set of undirected edges $E\subset V\times V$. The vector of $p$-variate random vector $\bm X=\bm\Theta+2\pi\bm K$ is an Unwrapped Normal graphical model with respect to $\mathcal{G}$ if $\bm \Theta\sim WN_p(\bm\mu,\bm{\Sigma})$, and 	$$\left(\bm \Sigma^{-1}\right)_{ij}=0\;\;\; \text{for all}\;\;\; \{i,j\}\notin E. $$
\end{definition}

The unwrapped variables $\bm X$ have, as a matter of fact,  a multivariate Gaussian distribution whose conditional independence structure can be described by a \emph{concentration graph} \citep{lauritzen1996}. However, the unwrapped variables are actually unobserved, as the binding vector $\bm K$ is unobserved. The concentration matrix driving the conditional independence structure of the Unwrapped graphical model depends on the parameters shared by the distributions of $\bm X$ and $\bm \Theta$. It then can be estimated using the observed angles.
As well known, the maximisation of the complete likelihood is infeasible for the wrapped Normal distribution, because of the infinite sums. However, its approximated version using $\bm k\in \{-1,0,1\}^p$ works quite well also for moderately large variances ($2\pi$).
Conversely, the approximation using $\bm k=\bm 0_p$ provides good estimates only for very small variances (for example, $0.0001$ when $p=5$).

Notice that, while the distribution of $\bm X$ factorizes with respect to $\mathcal{G}$, the distribution of $\bm \Theta$ can be written as a \emph{wrapped factorization}.  This is  the sum over $\bm k$ of the factorized distribution of $\bm X$. Indeed, assuming that $\bm X_A$ is independent of $\bm{X}_C$ given $\bm{X}_S$, with $C$ being a set of cardinality $c$ forming with $A$ and $S$ a partition of $\{1,\dots,p\}$,

\begin{align*}&f_{\bm{\Theta}}(\bm{\theta})=\sum_{\bm k_A\in\mathbb{Z}^q}\sum_{\bm k_C\in\mathbb{Z}^c}\sum_{\bm k_S\in\mathbb{Z}^s}f_{\bm X}(\bm \theta_A+2\pi\bm k_A,\bm\theta_C+2\pi\bm k_C,\bm \theta_S+2\pi\bm k_S)\\&=\sum_{\bm k_S\in\mathbb{Z}^s }\sum_{\bm k_A\in\mathbb{Z}^q}\sum_{\bm k_C\in\mathbb{Z}^c}f_{\bm X_A\mid \bm X_S}(\bm \theta_A+2\pi\bm k_A\mid \bm{\theta}_S+2\pi\bm k_S)\\&\cdot f_{\bm X_C\mid \bm X_S}(\bm \theta_C+2\pi\bm k_C\mid \bm{\theta}_S+2\pi\bm k_S)f_{\bm{X}_S}(\bm{\theta}_S+2\pi\bm k_S).\end{align*}

If $\bm X_A \ind \bm X_C \mid \bm X_S$, then $\bm \Theta_A \ind \bm \Theta_C \mid \bm X_S$ or equivalently $\bm \Theta_A \ind \bm \Theta_C \mid \bm\Theta_S, \bm K_S$. However, if $\bm X_A \ind \bm X_C \mid \bm X_S$, it is not implied that $\bm \Theta_A \ind \bm \Theta_C \mid \bm \Theta_S$.

When $\bm \Theta\sim WN_p(\bm \mu,\bm{\Sigma})$ and  $(\bm{\Sigma})_{ii}$ is small enough for any $i\in\{1,\ldots,p\}$, then the distribution of $\bm{\Theta}$ can be safely approximated by a $N_p(\bm{\mu},\bm{\Sigma})$ distribution. 
Moreover, if $(\bm \Sigma_{S})_{ii}$ is small enough for any  $i\in\{1,\ldots,s\}$, then
\begin{equation}\label{lim1}||f_{\bm \Theta_A\mid\bm{\Theta}_S}(\bm{\theta}_A\mid\bm \theta_S)-f_{\bm \Theta_A \mid\bm{X}_S}(\bm \theta_A\mid\bm{\theta}_S)||_{\infty}\to 0,\end{equation}
that is the distribution of $\bm \Theta_A\mid\bm{\Theta}_S$ can be approximated by the wrapped Normal distribution of $\bm\Theta_A\mid \bm X_S$. In such a situation, it holds that $w_S(\bm\theta_S,\bm k_s)$ goes to $0$ for any $\bm k_s\neq \bm 0_{s}$. 
Finally, whenever  $(\bm \Sigma_{AC\mid S})_{ii}$ is small enough for any  $i\in\{1,\ldots,q+c\}$, then
\begin{equation}\label{lim2}||f_{\bm \Theta_A\bm\Theta_C\mid\bm{\Theta}_S}(\bm{\theta}_A,\bm{\theta}_C\mid\bm \theta_S)-f_{\bm X_A \bm X_C\mid\bm{X}_S}(\bm \theta_A,\bm\theta_C\mid\bm{\theta}_S)||_{\infty}\to 0.\end{equation}
The condition for \eqref{lim1} is less stringent than the condition for \eqref{lim2}, as it admits nodes that are not separators, such as for instance singletons, to have larger variance. Simulations not reported here showed that the convergence \eqref{lim1} occurs faster than the convergence of the WN distribution to the Gaussian distribution. The graph of the unwrapped variables approximately describe the conditional independence structure of the angles, whenever the variances of the separators in $\mathcal{G}$ are small enough, that is \eqref{lim1}  or \eqref{lim2} hold.


\section{Inverse Stereographic Gaussian graphical models}\label{ISN}
A  vector of random angles $\bm \Theta=(\Theta_1\ldots\Theta_p)'$  has a Inverse Stereographic Normal distribution, $\bm \Theta\sim ISN_p(\bm{\mu},\bm{\Sigma})$, with $\bm \mu\in\mathbb{R}^p$ and $\bm \Sigma\in \textsf{Sym}_+^{p\times p}$,  if  the probability density function of   $\bm \Theta$ at $\bm\theta=(\theta_1\ldots\theta_p)'\in\mathbb{T}^p$ is
\begin{equation}\label{stereo}f_{\bm{\Theta}}(\bm{\theta})=\frac{1}{|2\pi\bm \Sigma|^{1/2}}\exp\left((\bm u-\bm{\mu})'\bm{\Sigma}^{-1}(\bm u-\bm{\mu})\right)\prod_{j=1}^p \frac{1}{1+\cos(\theta_j)},\end{equation}
where $\bm u=(u_1\ldots u_p )'$ with $u_j=\tan(\theta_j/2)$ being the \emph{stereographic projection} of the vector $(\cos(\theta_j)\;\;\sin(\theta_j))'$. For further details, see \cite{selvitella2019geometric}.

Given the stereographic projection $\bm U=\tan(\bm \Theta/2)$, if $\bm \Theta\sim ISN_p(\bm \mu,\bm \Sigma)$, then $\bm U\sim N_p(\bm \mu,\bm \Sigma)$ and the parameters of the Inverse Stereographic Normal distribution are the location and scale parameters of the distribution of $\bm U$ whose maximum likelihood estimators are the ordinary ones.

For $\bm{\Theta}\sim ISN_p(\bm{\mu},\bm{\Sigma})$, using partitions as in  $(\ref{partition})$, using \eqref{stereo}, 
it holds that
$$\bm\Theta_A\sim ISN_q(\bm \mu_A,\bm \Sigma_{AA}),\;\;\;and\;\;\;\bm\Theta_A\mid\bm\Theta_B\sim ISN_{q}(\bm{\mu_{A\mid B}},\bm{\Sigma}_{A\mid B}),$$
where $$\bm \mu_{A\mid B}=\bm{\mu}_A-\bm{\Sigma}_{AB}\bm{\Sigma}^{-1}_{BB}(\bm u_B-\bm{\mu}_B),\;\;\;\text{and}\;\;\;\;\bm{\Sigma}_{A\mid B}=\bm{\Sigma}_{AA}-\bm{\Sigma}_{AB}\bm\Sigma_{BB}^{-1}\bm{\Sigma}_{BA},$$
with $\bm u_B$  having  as its $j$th element the tangent of the halved $j$th entry of $\bm \theta_B$. Consequently,  if $\bm\Sigma_{AB}=\bm 0_{q\times (p-q)}$, then $\bm \Theta_A\ind \bm{\Theta}_B$. Moreover, the following result holds.
\begin{result}\label{stereo} Let $\bm \Theta\sim ISN_p(\bm \mu,\bm \Sigma)$. Consider $\bm\Theta_A, \bm\Theta_C,\bm\Theta_S$  where $A$, $C$ and $S$ are non-empty  disjoint subsets of $\{1, \ldots, p\}$ with respective cardinalities $q$, $c$ and $s$ such that $q+c+s \leq p$. Then $\bm \Theta_A\ind\bm \Theta_C\mid\bm{\Theta}_S$  iff 	
	$$
	\left ( \bm \Sigma_{ACS}\right)^{-1}_{AC} = \bm 0_{q \times c}.
	$$
	
\end{result}  
\begin{proof} See Appendix.
\end{proof}

When $\theta_j=-\pi$ for at least one $j\in\{1,\ldots,p\}$, $f_{\bm \Theta}(\bm \theta)$ has  removable singularities. A possible way to remove these singularities is to put at $0$ the values of the density at these points. For our purpose, it is convenient to replace the value of the density at singularity points with the value of the density when $\theta_j=-\pi+\epsilon_j$ with a non-zero, small enough $\epsilon_j$.  Using this latter convention, the resulting density is positive on $\mathbb{T}^p$, and the Hammersley-Clifford Theorem \citep{lauritzen1996} can be invoked to get the following definition.

\begin{definition} \textbf{\emph{(Inverse Stereographic Gaussian graphical model)}} Let $\mathcal{G}=(V,E)$ be an undirected graph with set of vertices $V=\{1,\ldots,p\}$ and set of undirected edges $E\subset V\times V$. The vector of random angles $\bm \Theta=(\Theta_1\ldots\Theta_p)'$ is an Inverse Stereographic Gaussian graphical model with respect to $\mathcal{G}$ if $\bm \Theta\sim ISN_p(\bm\mu,\bm{\Sigma})$ and 
	$$\left(\bm \Sigma^{-1}\right)_{ij}=0\;\;\; \text{for all}\;\;\; \{i,j\}\notin E. $$
	
\end{definition}

The assumption of strict positivity of $f_{\bm{\Theta}}$ implies in addition that $\bm\Theta_A\ind\bm{\Theta}_C\mid \bm{\Theta}_S $	whenever $S$ separates $A$ and $C$ in $\mathcal{G}$ for all disjoint subsets $A, C$ and $S$ of $V$ (global Markov property), that is  every path in $\mathcal{G}$ between two nodes $a\in A$ and $c\in C$ passes  through elements in $S$.    
Notice that because of the stereographic projection is a diffeomorphism,  the factorisation property for decomposable graphs directly follows by the respective property of the Gaussian graphical model (see \cite{lauritzen1996} for details).

A drawback of the ISN distribution is the lack of invariance under rotation. This issue suggests that the distribution should be used when the choice of the origin is not arbitrary. Alternatively, \cite{selvitella2019geometric} suggests to include the \emph{cut-point} as a further parameter in the model.

In many practical situations, the joint gaussianity assumption could be not plausible. In the Euclidean setting, a possible way to overcome this issue is to resort to more flexible distributions, such as the Nonparanormal distribution  \citep[see][]{liu2009nonpara}. 
In the case of circular random variables, we can define a toroidal counterpart of the Nonparanormal distribution as follows.

\begin{definition}\textbf{\emph{(Inverse Stereographic Nonparanormal distribution)}} We say that the  vector of random angles $\bm \Theta=(\Theta_1\dots\Theta_p)'$ has an Inverse Stereographic Nonparanormal distribution if there exists a set of functions $\bm h=\{h_1,\ldots ,h_p\}$ such that $\bm h(\bm U)=(h_1(U_1)\ldots h_p(U_p))'$, with $U_j=\tan(\Theta_j/2)$, satisfies $\bm h(\bm U)\sim N_p(\bm \mu,\bm{\Sigma})$.\end{definition}

Notice that, if $\bm h(\bm U)\sim N_p(\bm \mu,\bm{\Sigma})$, then $\bm U$ has a Nonparanormal distribution, $\bm U\sim NPN_p(\bm \mu,\bm{\Sigma},\bm h)$, and $\bm \Theta$ has an Inverse Stereographic Nonparanormal distribution,
$\bm{\Theta}\sim ISNPN_p(\bm{\mu},\bm{\Sigma},\bm h)$,  with density function

\begin{equation}\label{nonp}f_{\bm{\Theta}}(\bm{\theta})=\frac{1}{|2\pi\bm \Sigma|^{1/2}}\exp\left((\bm h(\bm u)-\bm{\mu})'\bm{\Sigma}^{-1}(\bm h(\bm u)-\bm{\mu})\right)\prod_{j=1}^p  \frac{|h'_j(u_j)|}{1+\cos(\theta_j)}.\end{equation}

As  in the Euclidean case,  to assure the identifiability of the above density function, we require that the transformation $h_j$ preserve means and variances for all $j$,  i.e.

$$\mathbb{E}[h_j(U_j)]=\mathbb{E}[U_j]=\mu_j,\;\;\;\mathbb{V}[h_j(U_j)]=\mathbb{V}[U_j]=(\bm \Sigma)_{jj}.$$
From  \eqref{nonp}, it can be easily seen that if $\bm \Theta\sim ISNPN_p(\bm \mu,\bm \Sigma,\bm h)$, then $\Theta_i\ind \Theta_j\mid \bm{\Theta}_{V/\{i,j\}}$, iff $(\bm{\Sigma^{-1}})_{ij}=0$. 
This enables us to define the following class of models. 

\begin{definition}\label{nonparanormal}\textbf{\emph{(Inverse Stereographic Nonparanormal graphical model)}} Let $\mathcal{G}=(V,E)$ be an undirected graph with set of vertices $V=\{1,\ldots,p\}$ and set of undirected edges $E\subset V\times V$. The vector of random angles $\bm \Theta=(\Theta_1\ldots\Theta_p)'$ is an  Inverse Stereographic Nonparanormal graphical model with respect to $\mathcal{G}$ if $\bm \Theta\sim ISNPN_p(\bm\mu,\bm{\Sigma})$ and 
$$\left(\bm \Sigma^{-1}\right)_{ij}=0\;\;\; \text{for all}\;\;\; \{i,j\}\notin E. $$
\end{definition}

Notice that, for the above class of models,  the global Markov property is implied by using the same convention on the singularity points as for the Inverse Stereographic Gaussian graphical model. The estimation of $\bm h$ and $(\bm{\mu},\bm{\Sigma})$ can be both carried out by using the same approach as in \cite{liu2009nonpara} over the realizations of $U_j$ and  $\widehat h(U_j)$, respectively.



\section{Illustrative examples: Using graphical models on the dihedral angles}\label{ex}

\begin{figure} 
	\centering
	\begin{tabular}{cc}
		\includegraphics[trim=1.5cm 0.cm .5cm .0cm, height =5.8cm]{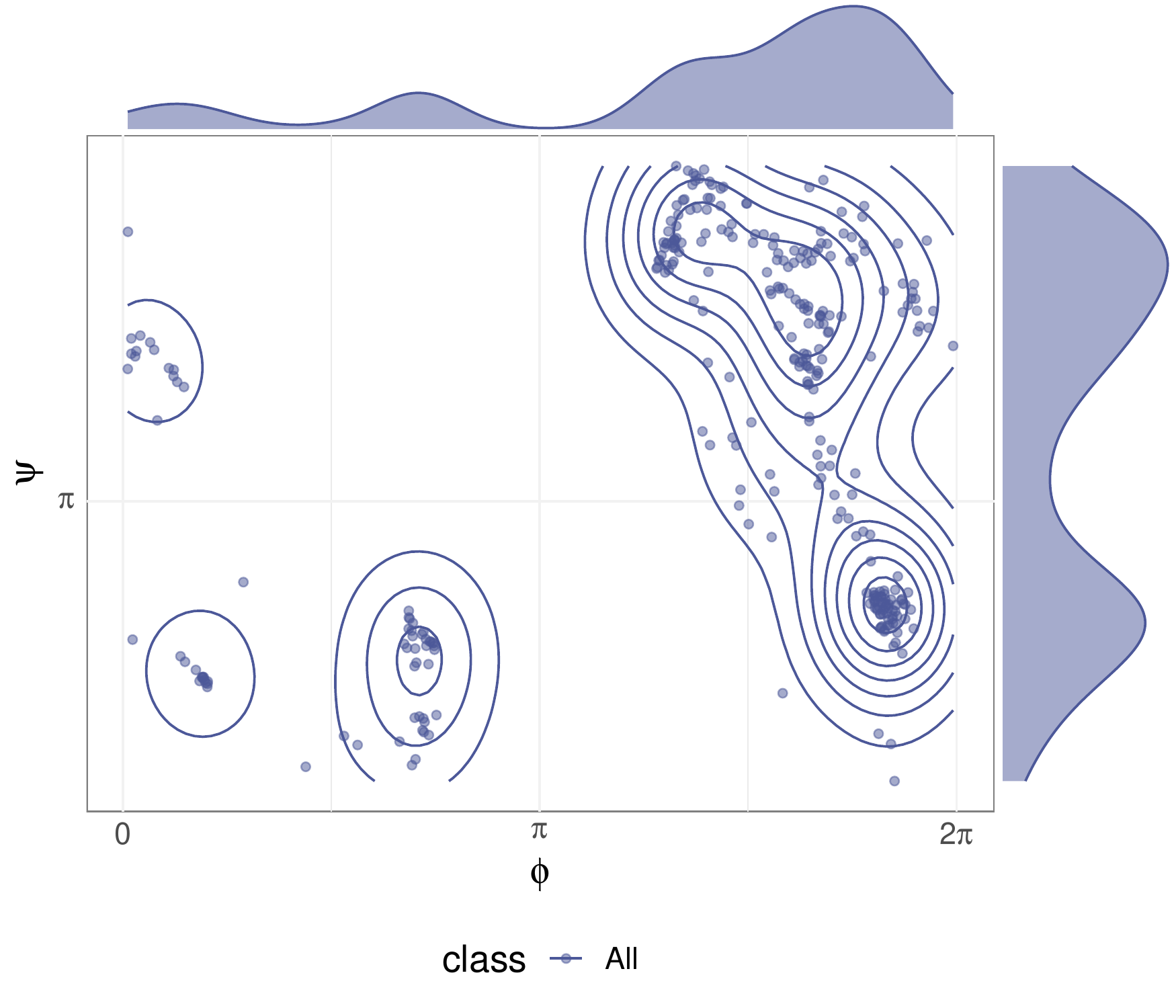} &	
		\includegraphics[trim=.3cm 0.cm .1cm 0.0cm,clip,  height =5.8cm]{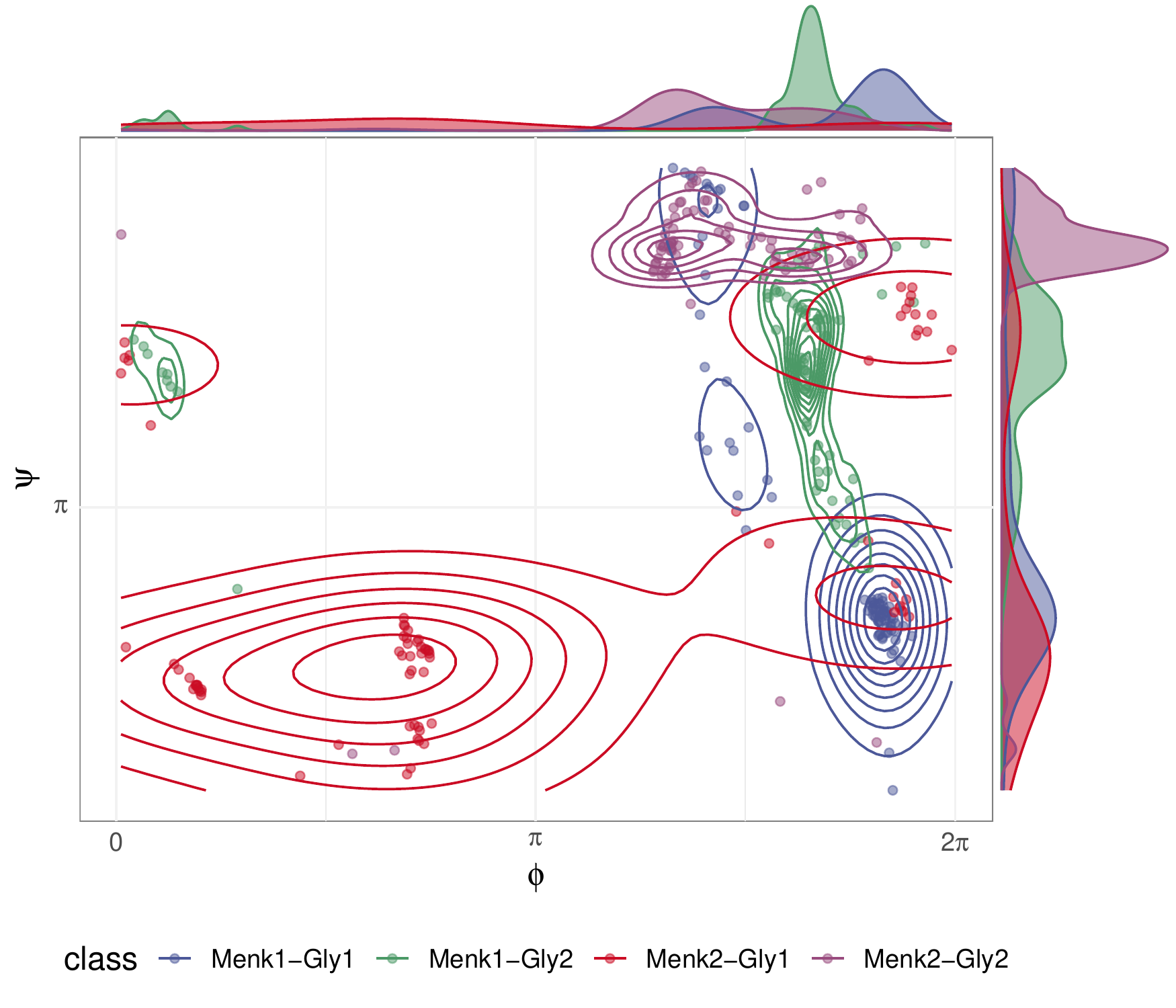}\\
		($a$) $\qquad \qquad \qquad$ & ($b$)
	\end{tabular}
	\caption{Ramachandran plot of two amino acids (Gly1 and Gly2), in Methionine-enkephalin (Menk) protein structures, measured in two experiments (Menk1 and Menk2). } \label{Rama}
\end{figure}

In structural bioinformatics, a crucial task consists in accurately predicting the three-dimensional structure of a protein.  Understanding the conditional independence structure linking the angles of a protein can provide substantial support to solve this task. 
The Ramachandran plot is the traditional way to visualise allowed regions for backbone dihedral angles $\phi$ and $\Psi$ of amino acid residues in protein structures.  Figure \ref{Rama} represents the Ramachandran plots of the dihedral angles $\phi$ and $\Psi$ for Glycine residues in Methionine-enkephalin (Menk) protein structures as described by \cite{marcotte2004}, where Menk was measured on two different experiments whose results are respectively called 1PLW and 1PLX (see Section \ref{secMenk} for some details).   

As we will see later, Menk contains two Glycine residues, called Gly1 and Gly2. Figure \ref{Rama} (b) is the Ramachandran plot which identifies Gly1 and Gly2 separately for both 1PLW and 1PLX. Figure \ref{Rama} (a) illustrates all the Glycine angles (for both the residues and both the experiments). Comparing the two plots, Figure \ref{Rama} (b) seems to suggest that the dependence between the two dihedral angles changes across the amino acids and the experiments and then, it could be interesting to study the dependence structure among all the angles in the protein structure.

Indeed, because the sequence of dihedral angles defines the backbone of a protein, the estimated conditional independence structure of dihedral angles could be useful for protein structure prediction. A convenient way to study the conditional independence structure is to adopt a graphical model where each node represents a specific dihedral angle of a specific amino acid in a protein.   
In the following, we demonstrate a few examples of the application of the proposed graphical models to evaluate the dependence structure of dihedral angles in two different proteins.

\subsection{The case of Methionine-enkephalin dihedral angles using WN graphical models}\label{secMenk}

Menk is a pentapeptide (five amino acids) endogenous opioid which is mainly found in the human central nervous system and gastrointestinal tract. Besides having an analgesic activity, the Menk is involved in the control of respiratory, cardiovascular, gastrointestinal functions, and neuroendocrine regulation. \cite{marcotte2004} carried out two interesting NMR experiments to measure the three-dimensional structure of Menk in a model membrane system similar to the natural environment of this pentapeptides. The main interest is in the effect of the membrane composition on the peptide conformation. They investigated two types of membranes,  zwitterionic (PC) bicelles and negatively charged bicelles (Bic/PG).
In each of the two experiments, they collected $n=80$ models of the Menk measured in fast-tumbling bicelles, using multidimensional 1H NMR. 
Both the data sets are available at the RCSB Protein Data Bank (\texttt{https://www.rcsb.org}), under the name of 1PLW, for the peptide conformation using zwitterionic PC bicelles, and 1PLX, when using Bic/PG bicelles.

We study these two sets of data separately, assuming a $p-$variate wrapped Normal distribution for the $p=8$ dihedral angles forming the protein
structure. For the estimation task, we consider an approximate maximum profile likelihood approach where the sum over $\bm k\in\mathbb{Z}^p$ has been replaced by the sum over $\bm k\in \{-1,0,1\}^p$, which is computationally intensive but still feasible. According to a simulation study we are not presenting here as out of the aim of the paper, this approximation works quite well whenever the elements of the diagonal matrix $\bm \Sigma$ are smaller than around $2 \pi$.  The mean vector is estimated via the circular sample mean.
For a first evaluation of the magnitude of the variances on the diagonal of $\bm \Sigma$, we apply component-wise Mardia variance estimator \citep{mardia2009directional}. 
As all these variance estimates are all relatively small, varying from $0.001$ to $0.036$ for 1PLX and  $0.003$ to $1.856$ for 1PLX. These estimates suggest that convergence to the Multivariate Normal is far to be reached, but the winding numbers can be truncated at $-1$ and $1$. 

\begin{figure} 
	\centering
	\begin{tabular}{cc}
		\includegraphics[trim=0.5cm 0.1cm .5cm .5cm, height =7.6cm]{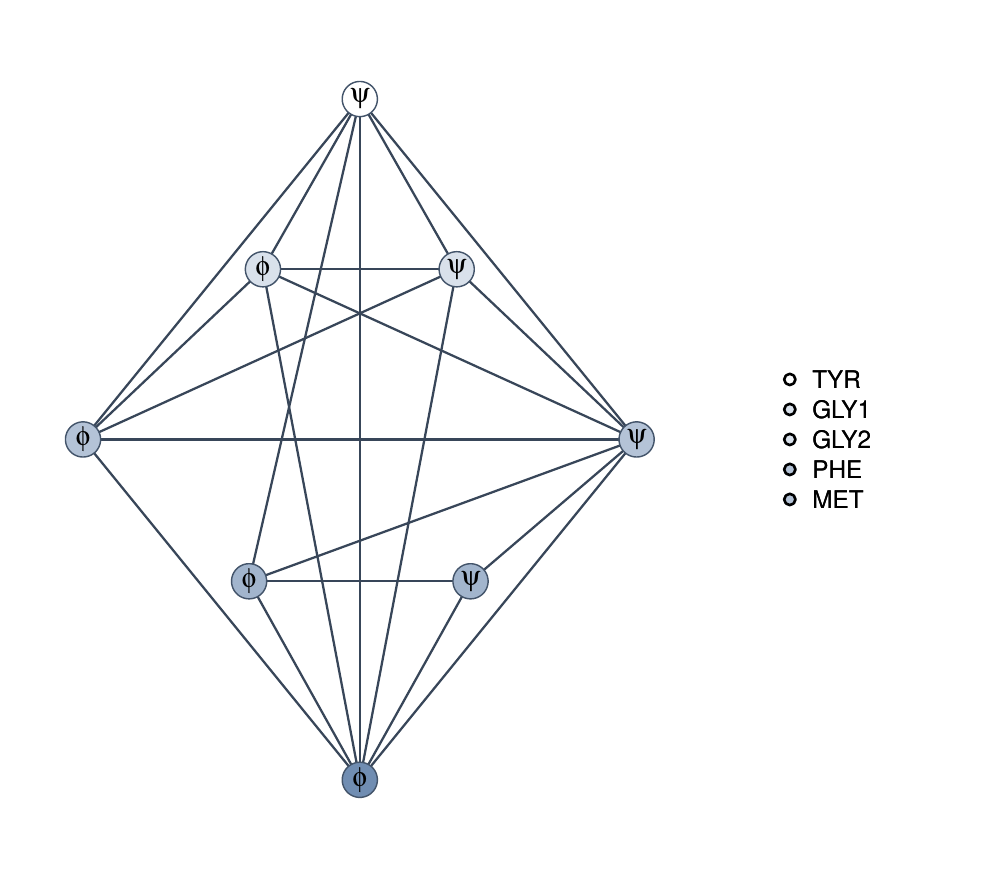} &	
		\includegraphics[trim=.5cm 0.1cm 3.5cm 0.5cm,clip,  height =7.6cm]{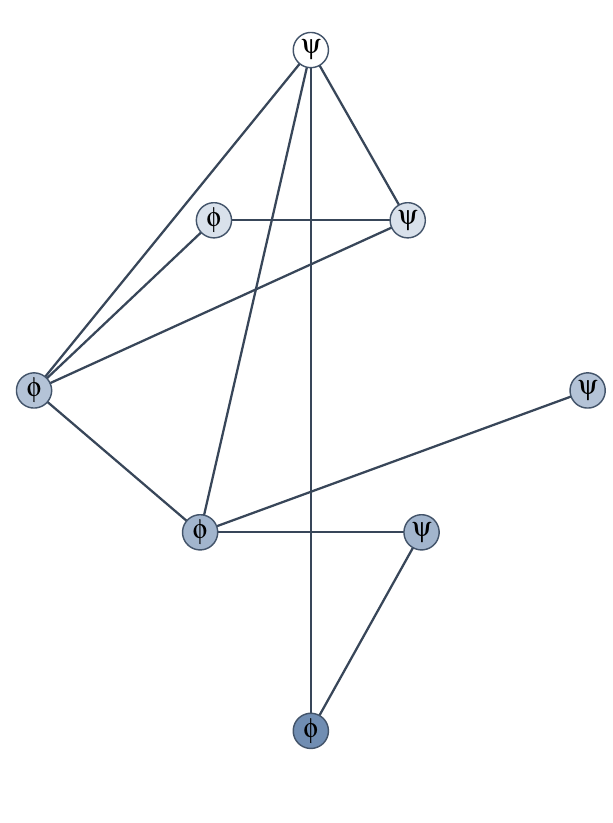}\\
		($a$) $\qquad \qquad \qquad$ & ($b$)
	\end{tabular}
	\caption{Selected unwrapped graphs for the angles of Metionine for 1PLX (a) and for 1PLW (b), with the overall significance level set at 0.05. }\label{meck_graph}
\end{figure}

Edge selection for the Unwrapped Normal graphical model (see Definition \ref{unwrapped}) is achieved by testing the off-diagonal elements of the variance-covariance matrix while controlling for the overall error rate for incorrect edge inclusion using Holm correction.
The unwrapped graphs are plotted in Figure \ref{meck_graph}. It is interesting to note that the graphs seem to confirm the effect of the membrane composition on
the peptide conformation. 
Specifically,  under the assumed model,  our results may be suggestive that Menk could adopt several conformations according to the membrane environments, not only in terms of means and variances of the dihedral angles as already observed by \cite{marcotte2004} but also on the dependence structure among them. 

\subsection{The reovirus p15 fusion-associated small transmembrane protein}

The second example concerns the reovirus fusion-associated small transmembrane protein, whose structure has been measured via NMR solution by \cite{read2015reovirus}. The data set is available at the RCSB Protein Data Bank under the name of 2MNS.
Reoviruses are a family of viruses that can affect the human gastrointestinal system and respiratory tract. The 2MNS protein has a chain structure with $22$ residue counts, with a total number of $40$ dihedral angles, measured on $n=50$ models from Baboon orthoreovirus by \cite{read2015reovirus}. 

When applying the inverse stereographic projection to the collected data, the Shapiro-Wilks test rejects Normality both on the univariate and the joint distributions.
However, the assumption of a Nonparanormal distribution is not rejected.
Accordingly, we assume that the vector of dihedral angles of the 2MNS protein has the ISNPN distribution in \eqref{nonp} whose transformation function $\bm h$ 
is estimated by using the same approach as in \cite{liu2009nonpara}. 

\begin{figure} 
	\centering
	\includegraphics[trim=.cm 0.cm 3.5cm .cm,clip,  width=13cm ]{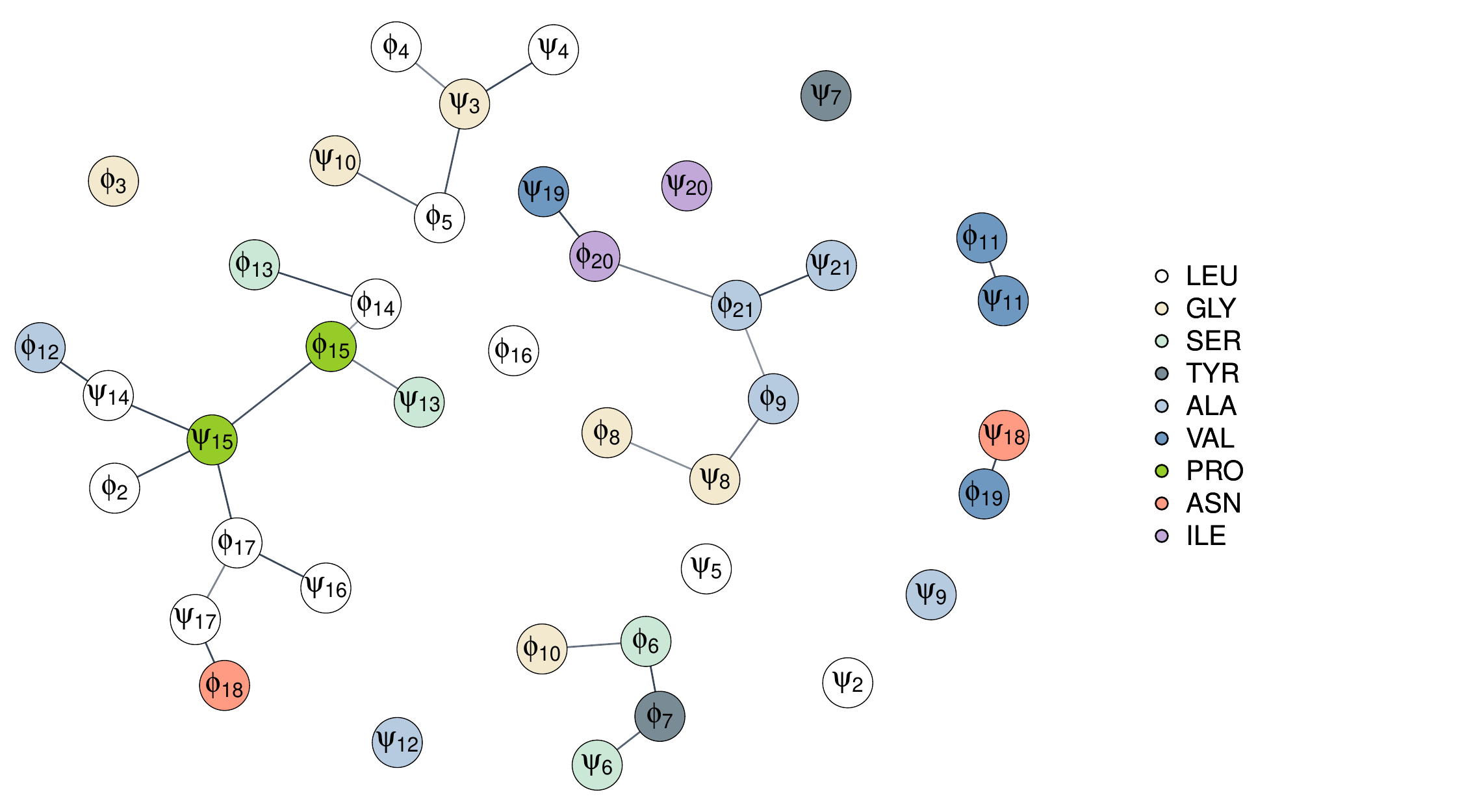}
	
	\vskip0.5cm	
	\caption{Selected graph for the 2MNS protein, where the subscript indicates the serial number of the component of the chain, while the colour indicates its type, reported in the legend.}\label{MNS_graph}
	
\end{figure}

Among the possible ways to learn the Inverse Stereographic Nonparanormal graphical model (see Definition \ref{nonparanormal}), we consider the adaptive lasso \citep{zou2006, kramer2009}, assuming sparsity. The tuning parameter of the estimation procedure is chosen via cross-validation. To improve stability, we repeatedly perform 5-fold cross-validation 50 times and take as stable the edges that were present at least the $50\%$ of the times.  

We observe a few interesting patterns, whatever choice of the stability threshold is made. As can be seen in Figure \ref{MNS_graph}, the dihedral angles appear less connected than the angles in Menk, with dependence mostly among angles that are neighbours in the primary structure.

\subsection{The structural characteristics of a mucin glycopeptide motif }

We now consider a study about the structural characteristics of a mucin glycopeptide motif. 
Data have been collected by \cite{coltart2002}, that investigated by NMR the secondary structure of a mucin glycopeptide derived from the N-terminal fragment of the cell surface glycoprotein CD43.
Mucin glycoproteins comprise one of the most relevant classes of cell surface molecules, serving a wide range of functions. Among many functions, \cite{coltart2002} pointed out the ability of mucin expression levels and architecture to serve as markers for the onset of disease, such as, for instance, upon carcinogenic transformation. For this reason, a good understanding of the structure of this protein could be relevant also for a cancer vaccine.  
The data are available at the RCSB Protein Data Bank, under the name of 1KYJ and include $n=59$ models of $p = 9$ dihedral angles. 

\begin{figure} 
	\centering
	\includegraphics[trim=0.5cm 0.1cm .5cm .6cm, height =7.4cm]{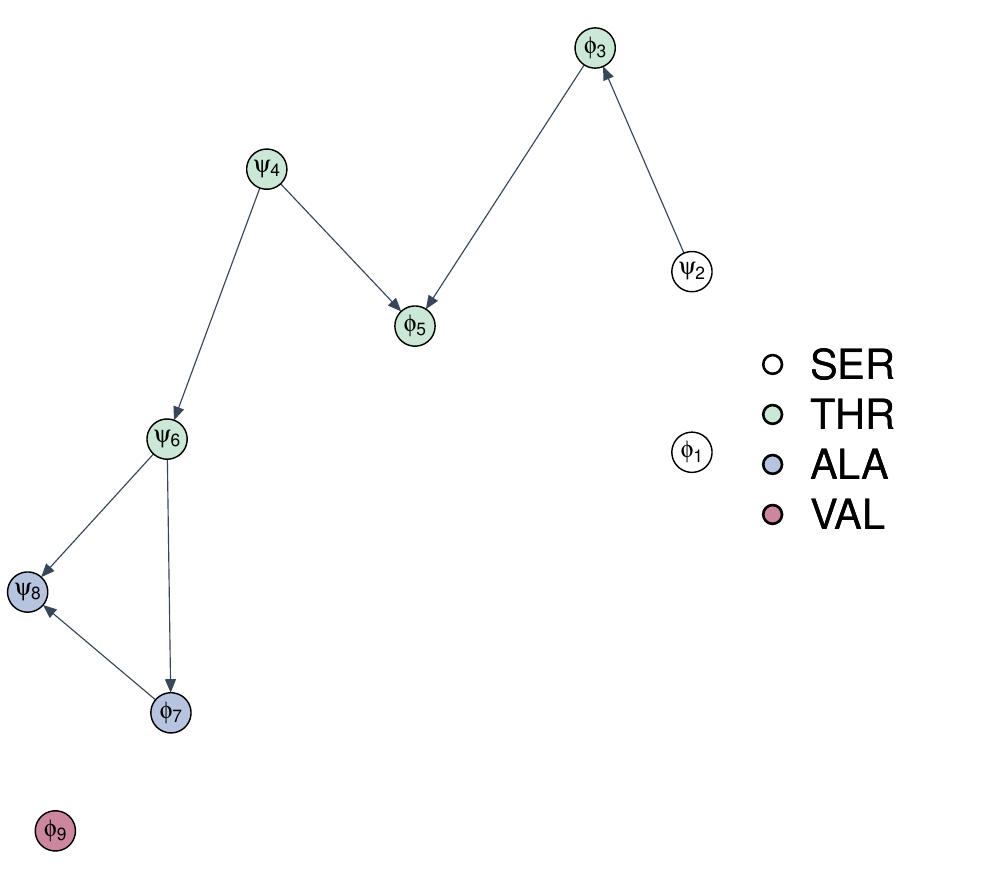} 
	\caption{Estimated DAG  for the angles of mucin glycopeptide, assuming the Conditional von Mises graphical model. }\label{mucine_graph}
\end{figure}

We assume a Conditional von Mises directed acyclic graphical models (see Definition \ref{condvM}), where the ordering is provided by the primary structure of the protein. The parameters estimates are obtained maximising the univariate conditional distributions. In particular, the conditional dependence parameters are obtaining maximising the profile likelihood, where the parameters $\mu_j$, $j=1, \ldots, p$, are estimated using the circular means. Edge selection has been reached using the likelihood ratio test for each distribution.  Figure \ref{mucine_graph} reports the selected graph. The dependence structure seems to follow the ribbon shape, also noted by \cite{coltart2002}.

\section{Conclusions}\label{concl}
Despite their potential, graphical models for angular variables seem to be under-studied. 
In this work, we aim to explore, in terms of conditional independence and graphical models, three probability distributions defined according to the main approaches used for specifying distributions on the $p-$dimensional torus: the intrinsic, the wrapping and the embedded approaches. 

We consider the multivariate von Mises distribution on the torus, the most popular distribution specified according to the intrinsic approach, whose closeness under conditioning of the univariate distributions makes it possible, under suitable conditions, the definition of a Bayesian network for angular variables. We call this model Conditional von Mises acyclic graphical model. It can be adopted whenever the ordering of the random angles is known. 
Within the wrapping approach, we considered the multivariate Wrapped Normal distribution. We analyse some properties of this distribution, and provide the definition of a class of undirected graphical models called the unwrapped Normal graphical model. 
Finally, we study the stereographic Normal distribution, based on the embedded approach, that leads to a further undirected graphical model along with a more flexible semi-parametric model. These models are called, respectively,  Inverse Stereographic Gaussian graphical model and Inverse Stereographic Nonparanormal graphical model. These models inherit all the properties of the corresponding models for variables defined on the real line. The possible extension to the definition of Bayesian networks is straightforward.

The interest in studying conditional independence structure of circular variables is shared by many scientific fields, and  has been strongly growing over the past years.
 In particular, in biochemistry, the structure of molecules such as proteins, DNA, and RNA  can be described in  terms of  angles.
Graphical models, whose nodes represent the angles within a molecule, may lead  to crucial contributions in biochemistry, and in
particular in the relevant problem of protein structure prediction.
In the paper, we show as the proposed graphical models can be applied for studying the local structure of the dihedral angles of some proteins.

\section*{Appendix} \textit{Proof of Result \ref{marginal}.} Let $B=\{1,\dots,p\}\setminus A$, then
\begin{align*}f_{\bm \Theta}(\bm\theta)&=\sum_{\bm k\in\mathbb{Z}^p}f_{\bm X}(\bm \theta+2\pi \bm k)\\&=\sum_{\bm k_A\in\mathbb{Z}^q}\sum_{\bm k_B\in\mathbb{Z}^{p-q}}|2\pi\bm{\Sigma}|^{-1/2}\exp\left\{-\frac{1}{2}\left(\begin{array}{c}\bm{\bm\theta}_A+2\pi\bm k_A\\\bm{\theta}_B+2\pi\bm k_B\end{array}\right)'\bm{\Sigma}^{-1}\left(\begin{array}{c}\bm{\theta}_A+2\pi\bm k_A\\\bm{\theta}_B+2\pi\bm k_B\end{array}\right)\right\}\\&=\sum_{\bm k_A\in\mathbb{Z}^q}\sum_{\bm k_B\in\mathbb{Z}^{p-q}}f_{\bm X}(\bm\theta_A+2\pi\bm k_A,\bm\theta_B+2\pi \bm k_B),\end{align*}
and
\begin{align*}f_{\bm \Theta_A}(\bm{\theta}_A)&=\int_{\mathbb{T}^{p-q}} \sum_{\bm k_A\in\mathbb{Z}^a}\sum_{\bm k_B\in\mathbb{Z}^b}f_{\bm X}(\bm \theta_A+2\pi\bm k_A,\bm \theta_B+2\pi\bm k_B)d\bm \theta_B\\
	&=\sum_{\bm k_A\in\mathbb{Z}^q}\sum_{\bm k_B\in\mathbb{Z}^{p-q}}\int_{\mathbb{T}^{p-q}}f_{\bm X}(\bm \theta_A+2\pi\bm k_A,\bm \theta_B+2\pi\bm k_B)d\bm \theta_B.\end{align*}
Consequently, by \emph{concatenation of integrals}

\begin{align*}f_{\bm \Theta_A}(\bm \theta_A)&=\sum_{\bm k_A\in\mathbb{Z}^q}\int_{ \mathbb{R}^{p-q}}f_{\bm X}(\bm{\theta}_A+2\pi\bm{k}_A,\bm {\theta}_B)d\bm\theta_B\\
	&=\sum_{\bm k_A\in\mathbb{Z}^q }f_{\bm X_A}(\bm{\theta}_A+2\pi\bm k_A), \end{align*}
where $\bm X_A\sim N(\bm{\mu}_A, \bm{\Sigma}_{AA})$. $\hfill\Box$
\vskip 1cm

\noindent \textit{Proof of Result \ref{conditionalX}.} As $\bm{X}_B=\bm{\Theta}_B+2\pi \bm{K}_B$,  the density of $\bm \Theta_{A},\bm K_A\mid \bm{\Theta}_{B},\bm K_B $ is 
\begin{align*}f_{\bm{\Theta}_A,\bm K_A\mid \bm{\Theta}_B, \bm K_B}(\bm\theta_A,\bm{k}_A\mid \bm{\theta}_B,\bm{k}_B)
	&=f_{\bm X_A\mid\bm X_B}(\bm \theta_A+2\pi\bm k_A\mid \bm \theta_B+2\pi\bm k_B ).\end{align*} 
Hence, as $\bm X_A\mid\bm X_B\sim N_q(\bm \mu_{A\mid B},\bm \Sigma_{A\mid B})$,  it holds that

$$f_{\bm \Theta_A\mid \bm{\Theta}_B,\bm K_B}(\bm \theta_A\mid \bm{\theta}_B,\bm k_B)=\sum_{\bm k_A\in\mathbb{Z}^q} f_{\bm X_A\mid\bm X_B}(\bm{\theta}_A+2\pi\bm k_A\mid\bm{\theta}_B+2\pi\bm k_B).$$
Therefore, $f_{\bm \Theta_A\mid \bm{\Theta}_B,\bm K_B}(\bm \theta_A\mid \bm{\theta}_B,\bm k_B)=f_{\bm \Theta_A}(\bm \theta_A)$ whenever $$f_{\bm X_A\mid\bm X_B}(\bm{\theta}_A+2\pi\bm k_A\mid\bm{\theta}_B+2\pi\bm k_B) = f_{\bm X_A}(\bm{\theta}_A+2\pi\bm k_A),$$ that is iff $\bm \Sigma_{AB}=\bm 0_{q \times (p-q)}$.$\hfill\Box$

\vskip 1cm

\noindent \textit{Proof of Result \ref{conditional}.} As a consequence of Result \ref{marginal}, 
$$
f_{\bm\Theta_A\mid \bm\Theta_S}(\bm\theta_A\mid \bm\theta_S) = 
\frac{\sum_{\bm k_A\in\mathbb{Z}^q}\sum_{\bm k_S\in\mathbb{Z}^s} 
	f_{\bm X_A \bm X_S}(\bm\theta_A + 2 \pi\bm k_A , \bm\theta_S + 2 \pi\bm k_S) }{ \sum_{\bm k_S\in\mathbb{Z}^s} f_{\bm X_S}(\bm\theta_S+2\pi \bm k_S)}.
$$ 
The result comes writing the joint density at the numerator as the product of the conditional density of $\bm X_A \mid \bm X_S$ and the marginal density of $\bm X_S$. $\hfill\Box$
\bigskip

\noindent\textit{Proof of Result \ref{stereo}.} The proof directly follows from Equation \eqref{stereo} and the conditional independence property of the Gaussian distribution.

\bibliographystyle{chicago}
\bibliography{mybib}
\end{document}